\documentclass[aps,prl,twocolumn,superscriptaddress]{revtex4-2}

\usepackage{amsmath}
\usepackage{amssymb}
\usepackage{braket}
\usepackage{graphicx}
\usepackage{color}
\usepackage{mathtools}
\usepackage{enumitem}
\usepackage{lipsum}
\usepackage{hyperref}

\graphicspath{{Figures/}}

\hypersetup{
	breaklinks=true,   
	colorlinks=true,   
	pdfusetitle=true,  
	citecolor=magenta,
	linkcolor=black
}

\begin{document}
	\title{Quantum Fluctuation Dynamics of Dispersive Superradiant Pulses \\ in a Hybrid Light-Matter System}
	\author{Kevin C. Stitely}
	\email{kevin.stitely@auckland.ac.nz}
	\affiliation{Dodd-Walls Centre for Photonic and Quantum Technologies, New Zealand}
	\affiliation{Department of Mathematics, University of Auckland, Auckland 1010, New Zealand}
	\affiliation{Department of Physics, University of Auckland, Auckland 1010, New Zealand}
	\author{Fabian Finger}
	\affiliation{Institute for Quantum Electronics, ETH Z\"{u}rich, 8093 Z\"{u}rich, Switzerland}
	\author{Rodrigo Rosa-Medina}
	\affiliation{Institute for Quantum Electronics, ETH Z\"{u}rich, 8093 Z\"{u}rich, Switzerland}
	\author{Francesco Ferri}
	\affiliation{Institute for Quantum Electronics, ETH Z\"{u}rich, 8093 Z\"{u}rich, Switzerland}
	\author{Tobias Donner}
	\affiliation{Institute for Quantum Electronics, ETH Z\"{u}rich, 8093 Z\"{u}rich, Switzerland}
	\author{Tilman Esslinger}
	\affiliation{Institute for Quantum Electronics, ETH Z\"{u}rich, 8093 Z\"{u}rich, Switzerland}
	\author{Scott Parkins}
	\affiliation{Dodd-Walls Centre for Photonic and Quantum Technologies, New Zealand}
	\affiliation{Department of Physics, University of Auckland, Auckland 1010, New Zealand}
	\author{Bernd Krauskopf}
	\affiliation{Dodd-Walls Centre for Photonic and Quantum Technologies, New Zealand}
	\affiliation{Department of Mathematics, University of Auckland, Auckland 1010, New Zealand}
	
	\date{\today}	
	
	\begin{abstract}
		We consider theoretically a driven-dissipative quantum many-body system consisting of an atomic ensemble in a single-mode optical cavity as described by the open Tavis-Cummings model. In this hybrid light-matter system the interplay between coherent and dissipative processes leads to superradiant pulses with a build-up of strong correlations, even for systems comprising hundreds to thousands of particles. A central feature of the mean-field dynamics is a self-reversal of two spin degrees of freedom due to an underlying time-reversal symmetry, which is broken by quantum fluctuations. We demonstrate a quench protocol that can maintain highly non-Gaussian states over long time scales. This general mechanism offers interesting possibilities for the generation and control of complex fluctuation patterns, as suggested for the improvement of quantum sensing protocols for dissipative spin-amplification.
	\end{abstract}
	
	\maketitle
	
	In recent years there has been considerable interest in driven-dissipative quantum many-body systems, which appear in fields ranging from atomic and optical physics \cite{langen_ultracold_2015}, to condensed matter \cite{carusotto_quantumfluid_2013} and quantum information theory \cite{harrington_engineered_2022}. The in- and outflux of energy, essential for applications, significantly modifies the dynamics that quantum systems admit. Examples exploiting the interplay between coherence and dissipation range from limit-cycle and time-crystalline behavior \cite{dogra_chiral_2019,Chiacchio_2019,buca_nonstationary_2019,kessler_timecrystal_2021}, superradiant oscillations \cite{zhiqiang_nonequilibrium_2017,stitely_superradiant_2020}, and chaos \cite{emary_chaos_2003,emary_quantum_2003,stitely_nonlinear_2020} to dissipation-induced and topological phases \cite{soriente_dissipation-induced_2018,dreon_topological_2022, diehl_topology_2011,ferri_emerging_2021,dreon_topological_2022}.
	Studies of many-body nonequilibrium quantum systems typically focus on mean-field behavior where quantum fluctuations are small. Moreover, treatments of fluctuations are usually concerned with steady-state effects, such as critical exponents in the vicinity of quantum phase transitions \cite{nagy_exponents_2011,oztop_exponents_2012,brennecke_fluctuations_2013}. Studies of the \emph{dynamics of quantum fluctuations} in driven-dissipative quantum systems have so far been limited to spin squeezing \cite{pezze_metrology_2018,ma_spinsqueezing_2011,cox_squeezing_2016,lewis_swan_squeezing_2018}. In these works, however, dissipation is deliberately mitigated as it is detrimental to entanglement.
	
\begin{figure}[b!]
		\centering
		\includegraphics[width=8cm]{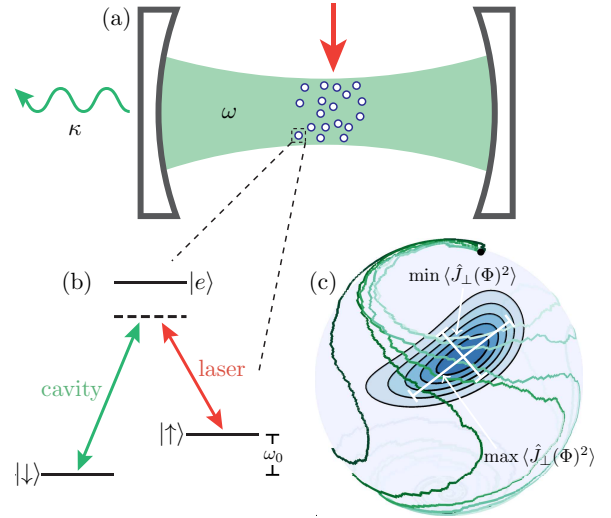}
		\caption{\label{fig:Schematic} (a) Schematic of the open Tavis--Cummings model. (b) Atomic energy level and coupling configuration. (c) Dispersive superradiant pulse illustrated on the atomic Bloch sphere by  six quantum trajectories and the spin $Q$-function; here, $\kappa = 1$, $\lambda = 0.5$, $N=200$, with initial state $\ket{\theta_0,\phi_0} = \ket{\pi/10,\pi/2}$.}
	\end{figure}

	In this work, we explore the fluctuation dynamics of an atomic ensemble coupled to a single-mode optical cavity via a driving laser field [Fig.~\ref{fig:Schematic}(a,b)]. This system, described by the dissipative Tavis-Cummings model \cite{tavis_TC_1968}, has been thoroughly investigated in its limiting cases where dissipation is either dominant, leading to superradiant decay \cite{dicke_coherence_1954}, or practically nonexistent, inducing spin-squeezing dynamics \cite{ma_spinsqueezing_2011,hu_vacuum_2017}. We study the unexplored regime between these two extremes, and show that the system supports a build-up of strong non-Gaussian fluctuations, which we term \emph{dispersive superradiant pulses} [Fig.~\ref{fig:Schematic}(c)]. Here, quantum fluctuations remain significant even as the system size increases towards a regime where one might na{\"i}vely expect the thermodynamic limit to take hold. Moreover, we show that the appearance of enhanced fluctuations leads to the breaking of an \emph{inversion-time reversal symmetry} of the thermodynamic limit. Finally, we demonstrate that quenching the effective cavity resonance and light-matter coupling can preserve highly non-Gaussian states over surprisingly long timescales.
	
	We consider $N$ atoms inside a single-mode optical cavity with effective resonance frequency $\omega$ and decay rate $\kappa$. The atoms are in a $\Lambda$ configuration with two (nominal) ground states, $\ket{\downarrow}$ and $\ket{\uparrow}$, and an excited state $\ket{e}$  [Fig.~\ref{fig:Schematic}(a,b)], and are driven with a cavity-assisted Raman transition: a laser couples to the transition $\ket{\uparrow}\leftrightarrow\ket{e}$, while $\ket{\downarrow}\leftrightarrow\ket{e}$ is coupled to the cavity mode. The atoms are fixed in place at alternate antinodes of the field and couple identically. If driving laser and cavity mode are far-detuned from atomic resonance, $\ket{e}$ can be adiabatically eliminated. The coupling scheme can then be seen as a two-photon transition between $\ket{\downarrow}$ and $\ket{\uparrow}$, facilitated by a laser and a cavity photon. 	
	The system is captured by the Tavis-Cummings Hamiltonian \cite{tavis_TC_1968} (with $\hbar = 1$)
	\begin{equation}\label{eqn:TC}
	\hat{H}_\text{TC} = \omega \hat{a}^\dagger\hat{a} + \omega_0\hat{J}_{z} + \frac{\lambda}{\sqrt{N}} \left( \hat{a}\hat{J}_{+} + \hat{a}^\dagger\hat{J}_{-} \right),
	\end{equation}
	where $\hat{a}$ is the annihilation operator of the cavity field mode with effective frequency $\omega$, $\hat{J}_{\pm,z}$ are collective spin-$N/2$ operators, $\omega_0$ is the effective energy level splitting, and $\lambda$ is the light-matter coupling strength. 
	The primary dissipative mechanism of this open quantum system is leakage of cavity photons, which we model with the Lindblad master equation
	\begin{equation}\label{eqn:TC_master}
	\frac{d\hat{\rho}}{dt} = -i[\hat{H}_\text{TC},\hat{\rho}] + \kappa\left( 2\hat{a}\hat{\rho}\hat{a}^\dagger - \hat{a}^\dagger\hat{a}\hat{\rho} - \hat{\rho}\hat{a}^\dagger\hat{a} \right),
	\end{equation}
	where $\hat{\rho}$ is the reduced density operator and $\kappa$  is the cavity decay rate. By adiabatically eliminating the cavity field mode (assuming $\sqrt{\omega^2 + \kappa^2}\gg\lambda,\omega_0$) \cite{masson_squeezing_2017,hu_vacuum_2017} and taking a rotating frame \footnote{\label{SM_footnote}See Supplemental Material at \textbf{LINK} for details on the semiclassical model, the second-order cumulant expansion, the Husimi spin $Q$-function, and the quench protocol.}, we arrive at a one-axis twisting Hamiltonian
	\begin{equation}\label{eqn:LMG}
	\hat{H} = -\frac{\xi\lambda^2}{N}\left( \frac{N^2}{4} - \hat{J}_{z}^2 \right),
	\end{equation}
	with the master equation
	\begin{equation}\label{eqn:LMG_master}
	\frac{d\hat{\rho}}{dt} = -i[\hat{H},\hat{\rho}] + \frac{\eta\lambda^2}{N}\left( 2\hat{J}_{-}\hat{\rho}\hat{J}_{+} - \hat{J}_{+}\hat{J}_{-}\hat{\rho} - \hat{\rho}\hat{J}_{+}\hat{J}_{-} \right).
	\end{equation}
	Here $\xi = \omega/(\kappa^2 + \omega^2)$,  $\eta = \kappa/(\kappa^2 + \omega^2)$, and $\hat{\rho}$ is now the reduced density operator of the atom-only system. Equation~(\ref{eqn:LMG}) describes the effective global-range spin-spin interactions within the ensemble, mediated by cavity photons. In Eq.~(\ref{eqn:LMG_master}), cavity dissipation gives rise to a collectively enhanced population inversion towards the spin state $\bigotimes_{j=1}^{N}\ket{\downarrow_j}$, i.e., Dicke superradiance \cite{dicke_coherence_1954,gross_superradiance_1982}. Since Eqs.~(\ref{eqn:LMG},\ref{eqn:LMG_master}) conserve the collective spin length, the dynamics can be represented as the evolution of a quasiprobability distribution on the surface of the Bloch sphere [Fig.~\ref{fig:Schematic}(c)], which we capture by the Husimi spin $Q$-function \footnote[1]{}. In the semiclassical limit $N\rightarrow\infty$, fluctuations become negligible (the quasiprobability distribution tends to a $\delta$ distribution) and the dynamics are described by the mean-field trajectories $\braket{\hat{J}_{x}(t)}$, $\braket{\hat{J}_{y}(t)}$, and $\braket{\hat{J}_{z}(t)}$, confined onto the Bloch sphere \footnote[1]{}.

\begin{figure}[t]
		\centering
		\includegraphics[width=8.6cm]{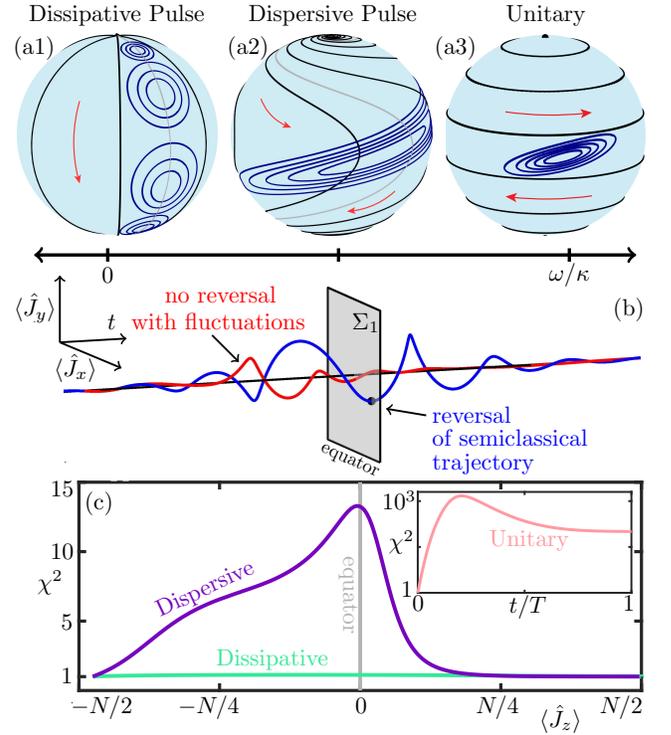}
		\caption{\label{fig:Expansion} (a) Semiclassical trajectories (black) and contours of the spin $Q$-function (blue) during the system's evolution on the Bloch sphere for dissipative (a1) and dispersive (a2) pulses initialized at $\ket{\theta_0,\phi_0} = \ket{\pi/10,\pi/2}$, and for the unitary case (a3) at $\ket{\pi/2,\pi/2}$. (b) Inversion-time reversal symmetry (blue) is broken by quantum fluctuations (red). (c) Deformation parameter $\chi^2$ as a function of the inversion $\braket{\hat{J}_{z}(t)}$ for $\omega = 0$ (dissipative), and for $\omega = 5$ (dispersive); inset: $\chi^2$ versus time for the unitary case with $\omega = 5$, $\kappa = 0$, and integration time $T = 800$. For all, $\lambda = 0.5$, $N=200$.}
	\end{figure}

	The open Tavis-Cummings model has been studied when the dynamics are either purely dissipative \cite{gross_superradiance_1982,andreev_superradiance_1980,angerer_superradiant_2018,clerk_spinamplifier_2022,argawal_master_1970,skribanowitz_observation_1973} or purely coherent (unitary) \cite{ma_spinsqueezing_2011,appel_mesoscopic_2009,wineland_spin_1992,wineland_squeezed_1994,norcia_cavity_2018}. The former case ($\omega/\kappa = 0$) reduces to the superradiance master equation \cite{gross_superradiance_1982,agarwal_master_1970}. Here, semiclassical trajectories move to the south pole along longitudinal lines [Fig.~\ref{fig:Expansion}(a1)]. Fluctuations increase uniformly in all spin directions due to the curvature of the Bloch sphere. The purely coherent case ($\omega/\kappa\rightarrow\infty$), on the contrary, has no dissipative term and exhibits spin squeezing dynamics \cite{ma_spinsqueezing_2011,pezze_metrology_2018,hu_vacuum_2017}. The semiclassical trajectories form latitudinal circles corresponding to Rabi oscillations [Fig.~\ref{fig:Expansion}(a3)], without emission into the cavity, and have an inversion-dependent Rabi frequency of $2\xi\lambda^2\braket{\hat{J}_{z}}/N$, see Eq.~(\ref{eqn:LMG}) --- leading to spin squeezing at the equator [Fig.~\ref{fig:Expansion}(a3)].
	
	In the intermediate range $0 < \omega/\kappa < \infty$, the interplay between superradiant-pulsed behavior via dissipation and Rabi oscillations from a cavity-atom detuning gives rise to rich dynamics of dispersive superradiant pulses.  Here, semiclassical trajectories unwind down the Bloch sphere and reverse direction at the equator [Fig.~\ref{fig:Expansion}(a2)], approaching the south pole. When the mean spin direction is at the equator, the (instantaneous) Rabi frequency is zero and becomes negative. This results in partial self-reversal of the trajectories, which stems from the \emph{inversion-time reversal symmetry} $\mathcal{T}_{1}: (t,\braket{\hat{J}_{z}}) \mapsto (-t,-\braket{\hat{J}_{z}})$ \footnote[1]{}. Hence, any trajectory that intersects the symmetry subspace ${\Sigma_1=  (\braket{\hat{J}_{x}},\braket{\hat{J}_{y}},\braket{\hat{J}_{z}})\in\mathbb{S}^2 | \braket{\hat{J}_{z}} = 0\}}$ is invariant under $\mathcal{T}_{1}$ \cite{devaney_reversible_1976,bandara_reversible_2021}. In other words, $\braket{\hat{J}_{x}}$ and $\braket{\hat{J}_{y}}$ undergo a reversal of their dynamics when the equator is crossed. 
	
	Since every trajectory tends towards the south pole of the Bloch sphere, any initially excited trajectory ($\braket{\hat{J}_{z}(0)} > 0$) is guaranteed to intersect $\Sigma_1$, and will thus be symmetric under inversion-time reversal. Unlike existing protocols for engineering time-reversal dynamics, which require external modification (flipping the sign of the Hamiltonian) \cite{garttner_reversal_2017,linnemann_reversal_2016,colombo_reversal_2022}, this system is intrinsically \emph{self-reversing} in the spin components $\braket{\hat{J}_{x}}$ and $\braket{\hat{J}_{y}}$ [Fig.~\ref{fig:Expansion}(b)]. Surprisingly, this is dissipation-induced: since the Rabi frequency at the equator is zero, motion through $\Sigma_1$ is purely dissipative. Note that simultaneous self-reversal of all spin components is impossible, as it would violate the uniqueness theorem for ordinary differential equations \cite{arnold_ODE_1992,glendinning_stability_1994}. Rather, only $\braket{\hat{J}_{x}}$ and $\braket{\hat{J}_{y}}$ self-reverse, while $\braket{\hat{J}_{z}}$ evolves monotonically to $\braket{\hat{J}_{z}}=-N/2$. 
	
	Simulation of Eq.~(\ref{eqn:LMG_master}) reveals that quantum fluctuations break inversion-time reversal symmetry [Fig.~\ref{fig:Expansion}(b)], connected with the appearance of highly non-Gaussian spin states during dispersive superradiant pulses [Figs.~\ref{fig:Schematic}(c) and~\ref{fig:Expansion}(a2)] with a large enhancement of fluctuations in one quadrature. The physical origin of the non-uniform state expansion can be understood by taking a quantum trajectories approach \cite{carmichael_stats2_2008,carmichael_open_1993}.  Fig.~\ref{fig:Schematic}(c)  shows six quantum trajectories initialized in a spin coherent state $\ket{\theta_0,\phi_0} = \ket{\pi/10,\pi/2}$. The positive divergence of the semiclassical equations in the northern portion of the Bloch sphere results in the amplification of the fluctuations of quantum trajectories during the initial stage of the pulse. This ``fanning out" of trajectories causes the enhancement of fluctuations in certain directions. Once the mean spin direction crosses the equator, the trajectories converge, so the overall level of fluctuations is reduced.
	
	To describe the non-uniform expansion, we introduce the \emph{deformation parameter}
	\begin{equation}\label{eqn:deformation}
	\chi^2 = \frac{\displaystyle\max_{\Phi}\braket{\Delta \hat{J}_{\perp}(\Phi)^2}}{\displaystyle\min_{\Phi}\braket{\Delta \hat{J}_{\perp}(\Phi)^2}} \geq 1,
	\end{equation}
	where $\braket{\Delta \hat{J}_{\perp}(\Phi)^2}$ is the spin variance in a direction orthogonal to the mean spin direction, parameterized by the angle $\Phi$, relative to the longitudinal direction. Figure~\ref{fig:Expansion}(c) shows $\chi^2$ during a purely dissipative and a dispersive pulse. In the dissipative case, the state expands uniformly [Fig.~\ref{fig:Expansion}(a1)] and $\chi^2\approx 1$ throughout. In contrast, the dispersive case has a clear initial rise in $\chi^2$, then peaks and quickly decreases as the system approaches its steady state ($\braket{\hat{J}_{z}}\rightarrow -N/2$) with uniform fluctuations. The (nonequilibrium) steady state is the spin coherent state at the south pole $\ket{\theta,\phi} = \ket{\pi,\phi}$ which is also a dark state of the jump operator $\hat{J}_{-}$, i.e., $\hat{J}_{-}\ket{\pi,\phi} = 0$. Figure~\ref{fig:Expansion}(c) also demonstrates that quantum fluctuations break inversion-time symmetry, since $\chi^2$ is not symmetric about $\braket{\hat{J}_{z}} = 0$.
	
	\begin{figure}[t]
		\centering
		\includegraphics[width=8.6cm]{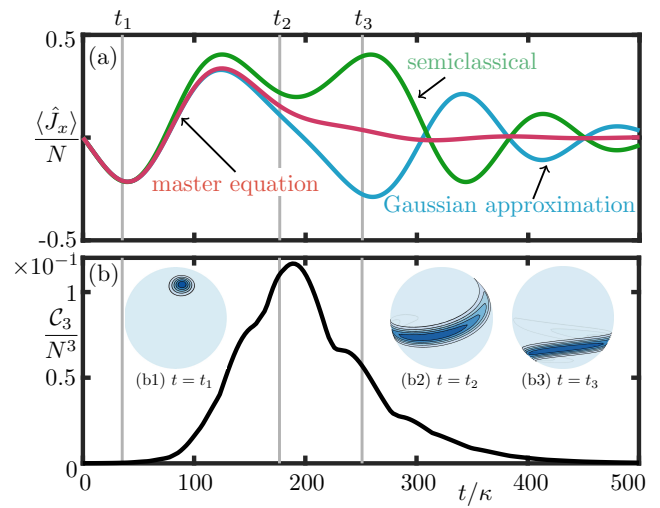}
		\caption{\label{fig:non_gaussian} Generation of non-Gaussian correlations in a dispersive superradiant pulse. (a) Semiclassical (green), second-order cumulant (blue), and master equation (red) computations of $\braket{\hat{J}_{x}(t)}$. (b) Total degree of correlation at third-order $\mathcal{C}_{3}(t)$. (b1)--(b3) Snapshots of the spin $Q$-function at different times of the pulse, with $\ket{\theta_0,\phi_0} = \ket{\pi/10,\pi/2}$ and $\kappa = 1$, $\omega = 5$, $\lambda = 0.5$, $N=200$.}
	\end{figure}
	
	We apply a cumulant expansion approach to further investigate these emerging non-Gaussian quantum states, as shown in Figs.~\ref{fig:Schematic}(c) and~\ref{fig:Expansion}(a2). Also known as connected correlation functions \cite{peskin_QFT_1995} or Ursell functions \cite{percus_correlation_1975}, cumulants quantify the effects of higher-order moments by subtracting out redundant information determined by moments of lower order. They have been used recently as the basis for a truncation method in a variety of contexts \cite{kubo_generalized_1962,kira_cluster_2008,sanchez_cumulant_2020,kirton_superradiant_2018,quach_superabsorption_2022,plankensteiner_quantumcumulantsjl_2022,huang_classical_2022}. 
	
	In typical light-matter interaction models, the equations of motion for moments at order $n$ depend on the moments at order $n+1$, thus, creating an infinite hierarchy of differential equations. In the $n$th cumulant expansion, one assumes that the cumulants at order $n+1$ vanish, thus closing the system of equations. The first-order expansion describes the semiclassical (mean-field) approximation, which is valid when the strength of fluctuations is small. The second-order cumulant expansion, with cumulants $\braket{\hat{J}_{j}\hat{J}_{k}}_{c} = \braket{\hat{J}_{j}\hat{J}_{k}} - \braket{\hat{J}_{j}}\braket{\hat{J}_{k}}$, $j,k\in\{x,y,z\}$, is valid for approximately Gaussian states and enables investigations of fluctuations, e.g., for spin squeezing. 
	The build-up of third-order cumulants
	\begin{align}\label{eqn:cumulant3}
	\braket{\hat{J}_{j}\hat{J}_{k}\hat{J}_{\ell}}_{c} =& \braket{\hat{J}_{j}\hat{J}_{k}\hat{J}_{\ell}} - \braket{\hat{J}_{j}}\braket{\hat{J}_{k}\hat{J}_\ell} - \braket{\hat{J}_{k}}\braket{\hat{J}_{j}\hat{J}_\ell} \nonumber \\ 
	&- \braket{\hat{J}_\ell}\braket{\hat{J}_{j}\hat{J}_{k}} + 2\braket{\hat{J}_{j}}\braket{\hat{J}_{k}}\braket{\hat{J}_{\ell}},
	\end{align}
	with $j,k,\ell\in\{x,y,z\}$, thus signals the appearance of non-Gaussianity. Since correlations shift between cumulants of different operator permutations dynamically, we define the \emph{total degree of correlation at third-order}
	\begin{equation}\label{eqn:C3}
	\mathcal{C}_{3}(t) = \sum_{j,k,\ell}|\braket{\hat{J}_{j}\hat{J}_{k}\hat{J}_{\ell}}_{c}|
	\end{equation}
	by summing over all possible cumulant permutations. 
	
	The time evolution of $\braket{J_x}$ and $\mathcal{C}_3$ during a dispersive superradiant pulse is shown in Figs.~\ref{fig:non_gaussian}(a,b).
	When the state is approximately coherent [Fig.~\ref{fig:non_gaussian}(b1)], $\mathcal{C}_{3}$ is small and the semiclassical and Gaussian approximations hold.
	Near the midpoint of the pulse, correlations build up and the state becomes highly non-Gaussian [Fig.~\ref{fig:non_gaussian}(b2)], leading to failure of the second-order cumulant expansion (Gaussian approximation) [Fig.~\ref{fig:non_gaussian}(a)].
	The final state is again a coherent spin state, for which $\mathcal{C}_3\to 0$. In the second-order cumulant expansion, inversion-time reversal symmetry is reflected by invariance under the transformation ${\mathcal{T}_{2}: (t,\braket{\hat{J}_{z}},\braket{\hat{J}_{z}\hat{J}_\ell})\mapsto (-t,-\braket{\hat{J}_{z}},-\braket{\hat{J}_{z}\hat{J}_\ell})},$
	where $\ell\in\{x,y\}.$ The corresponding symmetry subspace features the additional conditions $\braket{\hat{J}_{z}\hat{J}_{\ell}} = 0$. Contrary to the semiclassical case, where expectation factorization ensures $\braket{\hat{J}_{z}\hat{J}_{\ell}}=0$ if $\braket{\hat{J}_{z}}=0$, these are generally not satisfied. As a result, trajectories will generally not feature inversion-time reversal symmetry with quantum fluctuations.
	
	\begin{figure}[t]
		\centering
		\includegraphics[width=8.6cm]{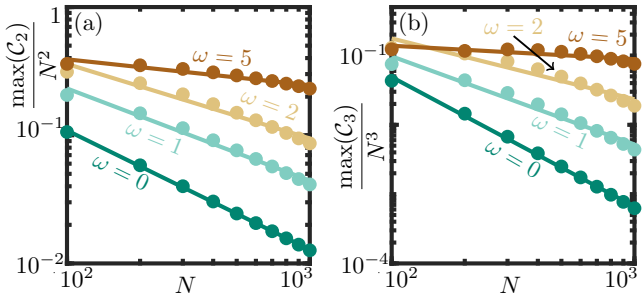}
		\caption{\label{fig:atom_number} Scaling of maximum correlations with $N$ at second order (a) and third order (b). Initial state is $\ket{\theta_0,\phi_0} = \ket{\pi/10,\pi/2}$ with $\kappa = 1$ and $\lambda = 0.5$.}
	\end{figure}
	
	So far, we studied the dissipative generation of non-Gaussian states for $N=200$. However, many cold atom experiments operate in the regime $N>10^3$ \cite{mivehar_cavity_2021}. Surprisingly, strong correlations created in the dispersive regime decay slowly as the atom number $N$ is increased. Figure~\ref{fig:atom_number}(a) shows how the maximum total degree of normalized correlations at order two, $\max(\mathcal{C}_{2})/N^2$, scales with $N$ as $\omega$ is varied. Linear fits over a logarithmic scale demonstrate an approximate power law. The purely dissipative case ($\omega  = 0$) has the largest gradient (magnitude), with correlations falling off rapidly (approximately $\propto1/N$), which is compatible with the behavior of coherent states. For larger $\omega$, both second- and third-order correlations, $\max(\mathcal{C}_{2})/N^2$ and $\max(\mathcal{C}_{3})/N^3$ [Fig.~\ref{fig:atom_number}] are maintained for significantly larger atom numbers, as the pulses become increasingly more dispersive. In this regime, we observe the formation of non-Gaussian states for systems with up to at least $N=1000$ particles. The onset of classical behavior, in the sense that higher-order cumulants vanish as $N$ increases, occurs for dissipative superradiant pulses well before dispersive pulses. Interestingly, this indicates that for typical atom numbers in cold atom experiments, a regime exists where even for large $N\gtrsim 10^3$ the full distribution of quantum fluctuations must be taken into account.
	
	\begin{figure}[t]
		\centering
		\includegraphics[width=8.6cm]{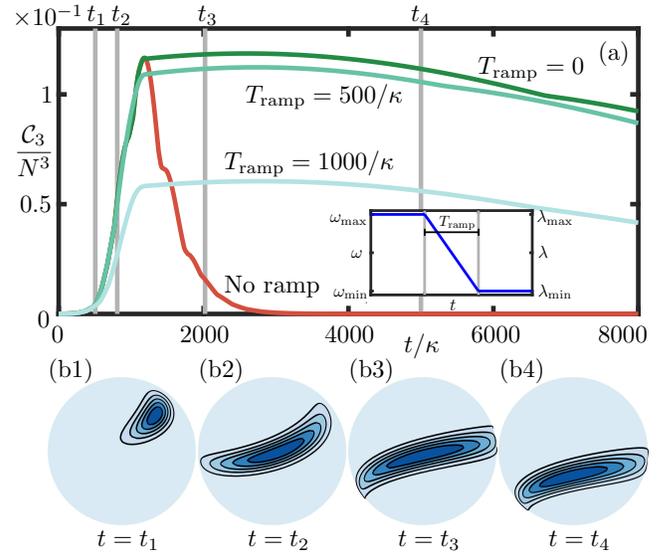}
		\caption{\label{fig:quench} Preserving non-Gaussian correlations with a quench protocol. (a) $\mathcal{C}_{3}(t)$ for no ramp (red), an instantaneous ramp (dark green), and for ramp times $T_\text{ramp} = 500/\kappa,1000/\kappa$ (lighter greens). The quench protocol is shown in the inset. (b1)--(b4) Snapshots of spin $Q$-function for $T_\text{ramp} = 500/\kappa$. Here, $\kappa = 1$, $\omega_\text{max} = 5$, $\omega_\text{min} = 0$, $\lambda_{\text{max}} = 0.2$, $\lambda_\text{min} = 0.02$, $N=200$, and the initial state is $\ket{\theta_0,\phi_0} = \ket{\pi/10,\pi/2}$.}
	\end{figure}
	
	In the dispersive pulses discussed so far, non-Gaussian states appear and disappear rapidly. As we now show, it is possible to form and subsequently maintain non-Gaussian states with a quenching protocol for the effective cavity resonance $\omega$ and coupling $\lambda$. When the mean spin direction is roughly at the equator of the Bloch sphere, so that non-Gaussian correlations are maximal, $\omega$ and $\lambda$ are ramped down. In this way, a dispersive pulse becomes dissipative and the evolution is slowed, thereby preserving non-Gaussian correlations are preserved [Fig.~\ref{fig:quench}]. To account for experimentally feasible, non-instantaneous quenches, we model the quenching protocol with a piecewise linear ramp in $\omega$ and $\lambda$ \footnote[1]{}. Initially, the pulse has a dispersive stage with $\omega = \omega_\text{max}$ and $\lambda = \lambda_\text{max}$. The initially excited atoms evolve as in Fig.~\ref{fig:non_gaussian} and begin to exhibit strong correlations, signalling the appearance of non-Gaussianity. When the correlations reach their peak near the midpoint of the pulse, $\omega$ and $\lambda$ are ramped down linearly over a time $T_\text{ramp}$ to a dissipative stage with $\omega = \omega_\text{min}$ and $\lambda = \lambda_\text{min}$. In a purely dissipative regime with $\omega_\text{min} = 0$, the non-Gaussian state formed over the initial dispersive stage of the pulse is ``frozen" and transported down the Bloch sphere as in Fig.~\ref{fig:Expansion}(a1). Since the speed is determined by $\lambda$, a quench can maintain the non-Gaussian states formed in the dispersive stage of the pulse over long periods of time. The preservation of correlations is monitored with $\mathcal{C}_{3}(t)$ in Fig.~\ref{fig:quench}(a), with the quenching protocol shown in the inset panel. With no ramping, the superradiant pulse remains dispersive, and there is a rapid build-up of correlations which then disappear quickly. For maintaining non-Gaussian correlations, the best-case scenario is an instantaneous quench with $T_\text{ramp} = 0$. In this case, we observe a rapid onset of correlations in the dispersive stage and then, in the dissipative stage, a decay that preserves third-order correlations over a long timescale (ensured by setting $\lambda_\text{min}/\lambda_\text{max}\ll 1$). For a more realistic quench with finite $T_\text{ramp}$ (the time taken for the driving laser to change frequency and power), we observe similar behavior, with the maximal degree of correlation only slightly below the instantaneous quench. The evolution of quantum states on the Bloch sphere for $T_\text{ramp} = 500/\kappa$ is shown in Fig.~\ref{fig:quench}(b1)--(b4), showcasing the dispersive stage, followed by a very slow decay of the state.
	
	In conclusion, we investigated the open Tavis-Cummings model in the dispersive regime between the well-explored limits of dominant dissipative and unitary processes. The interplay of both processes generates and sustains superradiant pulses with highly non-Gaussian quantum correlations. Importantly, enhanced fluctuations are largely preserved with increasing atom number, showing that the quantum-to-classical crossover can depend strongly on the dynamical behavior of fluctuations. Our results open up exciting possibilities for generation and control of specified non-Gaussian correlations in driven-dissipative many-body quantum systems. The non-Gaussian quantum states studied could be used in optimal state preparation in quantum metrology \cite{munoz-arias_phase_2022}. Owing to the strong state deformation, our protocol for creating dispersive superradiant pulses might be exploited to improve magnetic field sensing using superradiant spin-amplification \cite{clerk_spinamplifier_2022}.
	Its implementation with ultracold atoms in the dispersive regime of cavity QED \cite{zhiqiang_nonequilibrium_2017,ferri_emerging_2021} has several advantages. Since cavity detuning and coupling strength are determined by the properties of the drive laser, the parameters we considered can indeed be adjusted dynamically during the experiment. Moreover, it is possible to non-destructively measure the cavity output field, enabling real-time readout of the spin inversion, which is crucial for optimal superradiant spin amplification.
	\begin{acknowledgements}
		We acknowledge funding from the Swiss National Science Foundation: project numbers IZBRZ2, 186312, 182650, and 212168 and NCCR QSIT, from EU Horizon2020: ERC advanced grant TransQ (project number 742579).
	\end{acknowledgements}
	\bibliography{DispFluctRefs.bib}
	
\end{document}


\title{\Large{Supplemental Material:} \\\large{Quantum Fluctuation Dynamics of Dispersive Superradiant Pulses \\ in a Hybrid Light-Matter System}}
	
	\author{Kevin C. Stitely}
	\affiliation{Dodd-Walls Centre for Photonic and Quantum Technologies, New Zealand}
	\affiliation{Department of Mathematics, University of Auckland, Auckland 1010, New Zealand}
	\affiliation{Department of Physics, University of Auckland, Auckland 1010, New Zealand}
	\author{Fabian Finger}
	\affiliation{Institute for Quantum Electronics, ETH Z\"{u}rich, 8093 Z\"{u}rich, Switzerland}
	\author{Rodrigo Rosa-Medina}
	\affiliation{Institute for Quantum Electronics, ETH Z\"{u}rich, 8093 Z\"{u}rich, Switzerland}
	\author{Francesco Ferri}
	\affiliation{Institute for Quantum Electronics, ETH Z\"{u}rich, 8093 Z\"{u}rich, Switzerland}
	\author{Tobias Donner}
	\affiliation{Institute for Quantum Electronics, ETH Z\"{u}rich, 8093 Z\"{u}rich, Switzerland}
	\author{Tilman Esslinger}
	\affiliation{Institute for Quantum Electronics, ETH Z\"{u}rich, 8093 Z\"{u}rich, Switzerland}
	\author{Scott Parkins}
	\affiliation{Dodd-Walls Centre for Photonic and Quantum Technologies, New Zealand}
	\affiliation{Department of Physics, University of Auckland, Auckland 1010, New Zealand}
	\author{Bernd Krauskopf}
	\affiliation{Dodd-Walls Centre for Photonic and Quantum Technologies, New Zealand}
	\affiliation{Department of Mathematics, University of Auckland, Auckland 1010, New Zealand}
	
	\maketitle
	
	\section{Semiclassical Model}
	
	\noindent To recall, we study the Hamiltonian 
	\begin{equation}
		\hat{H} = -\frac{\xi\lambda^2}{N}\left( \frac{N^2}{4} - \hat{J}_{z}^2 \right),
	\end{equation}
	with the master equation
	\begin{equation}\label{eqn:master_equation}
		\frac{d\hat{\rho}}{dt} = -i[\hat{H},\hat{\rho}] + \frac{\eta \lambda^2}{N}\left( 2\hat{J}_{-}\hat{\rho}\hat{J}_{+} - \hat{J}_{+}\hat{J}_{-}\hat{\rho} - \hat{\rho}\hat{J}_{+}\hat{J}_{-} \right).
	\end{equation}
	With
	\begin{align}
		\beta &= \frac{\braket{\hat{J}_{-}}}{N} \in\mathbb{C}, \\
		\gamma &= \frac{\braket{\hat{J}_z}}{N} \in \mathbb{R},
	\end{align}
	the semiclassical approximation asserts that the expectations of operator products factorize, i.e., $\braket{\hat{J}_{j}\hat{J}_{k}} \approx \braket{\hat{J}_j}\braket{\hat{J}_{k}}$, $j,k\in\{x,y,z\}$. Thus, a closed set of ordinary differential equations for $\beta$ and $\gamma$ can be obtained by taking the trace over Eq.~(\ref{eqn:master_equation}), to obtain
	\begin{align}
		\frac{1}{N}\mathrm{Tr}\left( \hat{J}_{-}\frac{d\hat{\rho}}{dt} \right)=\frac{d\beta}{dt} &= 2(\eta - i\xi)\lambda^2 \beta\gamma, \label{eqn:semi1} \\
		\frac{1}{N}\mathrm{Tr}\left( \hat{J}_{z}\frac{d\hat{\rho}}{dt} \right)=\frac{d\gamma}{dt} &= -2\eta\lambda^2 \beta^* \beta. \label{eqn:semi2}
	\end{align}
	Here, spin conservation is expressed by the fact that $|\beta|^2 + \gamma^2$ is constant, confining $(\beta,\gamma)$ onto the Bloch sphere and making the dynamics essentially two-dimensional. The system is naturally represented in spherical coordinates,
	\begin{align}
		\beta &= r\sin(\theta)e^{-i\phi}, \\
		\gamma &= r\cos(\theta),
	\end{align}
	with radius $r \geq 0$, polar angle $\theta \in [0,\pi]$, and azimuthal angle $\phi \in [0,2\pi)$
	which leads to the set of ordinary differential equations
	\begin{align}
		\frac{d\theta}{dt} &= \eta\lambda^2 \sin(\theta), \\ 
		\frac{d\phi}{dt} &= \xi\lambda^2 \cos(\theta),
	\end{align}
	for the angles $\theta(t)$ and $\phi(t)$.
	Here the equation for $r$ has been eliminated since $dr/dt = 0$ due to spin conservation, and we have taken the radius to take on its maximal value of $r=1/2$. In this representation the system's $U(1)$ symmetry is represented by invariance under the transformation $\mathcal{R}(\varphi):\phi \mapsto \phi + \varphi$. Equations (9) and (10) were integrated numerically to compute the semiclassical trajectories on the Bloch sphere in Fig.~2(a1)--(a3), and the temporal traces in Fig.~2(b) and Fig.~3(a) in the main text.
	
	From the imaginary part of Eq.~(\ref{eqn:semi1}), it can be seen that the inversion-dependent Rabi frequency, which is characteristic of one-axis twisting dynamics, is $\Omega = 2\xi\lambda^2\gamma$; it becomes time-dependent when $\gamma$ decays via cavity dissipation.
	
	\section{Cumulant Expansion}
	
	In the semiclassical approximation above, the effects of quantum fluctuations are neglected in favor of a closed set of differential equations for the mean fields. Another way of phrasing this is: the second-order cumulants $\braket{\hat{J}_{j}\hat{J}_{k}}_{c}$, $j,k\in\{x,y,z\}$ vanish when the number of atoms $N$ becomes sufficiently large. 
	
	In general, the cumulants at order $n$ for the operators $\{\hat{A}_{1},\hat{A}_{2},\cdots,\hat{A}_{n}\}$ can be written as \cite{kubo_generalized_1962,plankensteiner_quantumcumulantsjl_2022}
	\begin{equation}\label{eqn:cumulants}
		\bigg\langle \prod_{i=1}^{n}\hat{A}_{i} \bigg\rangle_{c} = \sum_{p\in\mathcal{P}} (|p|-1)!(-1)^{|p|-1}\prod_{B\in p}^{|p|}\bigg\langle \prod_{m \in B}^{|B|} \hat{A}_{m} \bigg\rangle,
	\end{equation}
	where $\mathcal{P}$ is the set of all partitions of $\{1,2,\cdots,n\}$, $|p|$ is the length of the partition $p$, and $B$ is a block of the partition. For instance, $\{\{1,2\},3\}$ is a partition of $\{1,2,3\}$, where $\{1,2\}$ and $\{3\}$ are the blocks of the partition, and the length of the partition is two (the number of blocks). Equation~(\ref{eqn:cumulants}) reduces to the covariances for $n=2$ and Eq.~(7) of the main text for $n=3$. Importantly, the cumulants at order $n$ depend only on moments of order $n$ and below. 
	
	\subsection{Second-order Cumulant Expansion Model}
	
	The equations of motion for an arbitrary operator $\hat{A}$ under the evolution of the master equation~(\ref{eqn:master_equation}) come in the form
	\begin{equation}
	\frac{d}{dt}\braket{\hat{A}} = \frac{i\xi\lambda^2}{N}\braket{[\hat{A},\hat{J}_{x}^2 + \hat{J}_{y}^2]} + \frac{\eta\lambda^2}{N} \left( \braket{[\hat{J}_{+},\hat{A}]\hat{J}_{-}} + \braket{\hat{J}_{+}[\hat{A},\hat{J}_{-}]} \right).
	\end{equation}
	The equations for the first-order moments, taking $\hat{A}=\hat{J}_{j}$ for $j\in\{x,y,z\}$, yields
	\begin{align}
	\frac{d}{dt}\braket{\hat{J}_{x}} &= -\frac{\xi\lambda^2}{N}\bigg( \braket{\hat{J}_{z}\hat{J}_{y}} + \braket{\hat{J}_y \hat{J}_z} \bigg) + \frac{\eta\lambda^2}{N}\bigg( \braket{\hat{J}_z \hat{J}_-} + \braket{\hat{J}_+ \hat{J}_z} \bigg), \\
	\frac{d}{dt}\braket{\hat{J}_y} &= \frac{\xi\lambda^2}{N}\bigg( \braket{\hat{J}_z \hat{J}_x} + \braket{\hat{J}_x \hat{J}_z} \bigg) + \frac{\eta\lambda^2}{N}\bigg( i\braket{\hat{J}_z \hat{J}_-} - i\braket{\hat{J}_+ \hat{J}_z} \bigg), \\
	\frac{d}{dt}\braket{\hat{J}_z} &= -\frac{\eta\lambda^2}{N}\bigg( i\braket{\hat{J}_y \hat{J}_-} + \braket{\hat{J}_x \hat{J}_-} - i\braket{\hat{J}_+ \hat{J}_y} + \braket{\hat{J}_+ \hat{J}_x} \bigg).
	\end{align}
	The equations for the second-order moments ($\hat{A}=\hat{J}_{j}\hat{J}_{k}$, $j,k\in\{x,y,z\}$) are 
	\begingroup
	\allowdisplaybreaks
	\begin{align}
	\frac{d}{dt}\braket{\hat{J}_{x}^2} ={}& -\frac{\xi\lambda^2}{N}\bigg( \braket{\hat{J}_x \hat{J}_z \hat{J}_y} + \braket{\hat{J}_z \hat{J}_x \hat{J}_y} + \braket{\hat{J}_y \hat{J}_x \hat{J}_z} + \braket{\hat{J}_y \hat{J}_z \hat{J}_x} \bigg) \nonumber \\
	&+ \frac{\eta\lambda^2}{N}\bigg( \braket{\hat{J}_z \hat{J}_x \hat{J}_-} + \braket{\hat{J}_x \hat{J}_z \hat{J}_-} + \braket{\hat{J}_+ \hat{J}_z \hat{J}_x} + \braket{\hat{J}_+ \hat{J}_x \hat{J}_z} \bigg), \\
	\frac{d}{dt}\braket{\hat{J}_x \hat{J}_y} ={}& \frac{\xi\lambda^2}{N}\bigg( \braket{\hat{J}_x \hat{J}_z \hat{J}_x} + \braket{\hat{J}_{x}^2 \hat{J}_z} - \braket{\hat{J}_z \hat{J}_{y}^2} - \braket{\hat{J}_y \hat{J}_z \hat{J}_y} \bigg) \nonumber \\
	&+ \frac{\eta\lambda^2}{N}\bigg( \braket{\hat{J}_z \hat{J}_y \hat{J}_-} + i\braket{\hat{J}_x \hat{J}_z \hat{J}_-} + \braket{\hat{J}_+ \hat{J}_z \hat{J}_y} - i\braket{\hat{J}_+ \hat{J}_x \hat{J}_z} \bigg), \\
	\frac{d}{dt}\braket{\hat{J}_{x}\hat{J}_{z}} ={}& \frac{\xi\lambda^2}{N}\bigg( -\braket{\hat{J}_x \hat{J}_y \hat{J}_x} - \braket{\hat{J}_{z}^2 \hat{J}_y} + \braket{\hat{J}_y \hat{J}_{x}^2} - \braket{\hat{J}_y \hat{J}_{z}^2} \bigg) \nonumber \\
	&+ \frac{\eta\lambda^2}{N}\bigg( \braket{\hat{J}_{z}^2 \hat{J}_-} - i\braket{\hat{J}_x \hat{J}_y \hat{J}_-} - \braket{\hat{J}_{x}^2 \hat{J}_-} + \braket{\hat{J}_+ \hat{J}_{z}^2} + i\braket{\hat{J}_+ \hat{J}_x \hat{J}_y} - \braket{\hat{J}_+ \hat{J}_{x}^2} \bigg), \\
	\frac{d}{dt}\braket{\hat{J}_{y}^2} ={}& \frac{\xi\lambda^2}{N} \bigg( \braket{\hat{J}_x \hat{J}_z \hat{J}_y} + \braket{\hat{J}_z \hat{J}_x \hat{J}_y} + \braket{\hat{J}_y \hat{J}_x \hat{J}_z} + \braket{\hat{J}_y \hat{J}_z \hat{J}_x} \bigg) \nonumber \\
	&+ \frac{\eta\lambda^2}{N}\bigg( i\braket{\hat{J}_z \hat{J}_y \hat{J}_-} + i\braket{\hat{J}_y \hat{J}_z \hat{J}_-} - i\braket{\hat{J}_+ \hat{J}_z \hat{J}_y} - i\braket{\hat{J}_+ \hat{J}_y \hat{J}_z} \bigg), \\
	\frac{d}{dt}\braket{\hat{J}_{y} \hat{J}_{z}} ={}& \frac{\xi\lambda^2}{N} \bigg( \braket{\hat{J}_{z}^2 \hat{J}_{x}} - \braket{\hat{J}_x \hat{J}_{y}^2} + \braket{\hat{J}_x \hat{J}_{z}^2} + \braket{\hat{J}_y \hat{J}_x \hat{J}_y} \bigg) \nonumber \\
	&+ \frac{\eta\lambda^2}{N}\bigg( i\braket{\hat{J}_{z}^2 \hat{J}_{-}} - i\braket{\hat{J}_{y}^2 \hat{J}_{-}} - \braket{\hat{J}_y \hat{J}_x \hat{J}_-} - i\braket{\hat{J}_+ \hat{J}_{z}^2} + i\braket{\hat{J}_+ \hat{J}_{y}^2} - \braket{\hat{J}_+ \hat{J}_y \hat{J}_x} \bigg), \\
	\frac{d}{dt}\braket{\hat{J}_{z}^2} ={}& \frac{\eta\lambda^2}{N}\bigg(-i\braket{\hat{J}_y \hat{J}_z \hat{J}_-} - \braket{\hat{J}_x \hat{J}_z \hat{J}_-} - i\braket{\hat{J}_z \hat{J}_y \hat{J}_-} - \braket{\hat{J}_z \hat{J}_x \hat{J}_-} \\ 
	&+i\braket{\hat{J}_+ \hat{J}_y \hat{J}_z} -\braket{\hat{J}_+ \hat{J}_x \hat{J}_z} + i\braket{\hat{J}_+ \hat{J}_z \hat{J}_y} - \braket{\hat{J}_+ \hat{J}_z \hat{J}_x} \bigg).
	\end{align}
	\endgroup
	
	For truncation at third order, $\braket{\hat{J}_{j}\hat{J}_{k}\hat{J}_{\ell}}_{c} = 0$, the third-order moments are written in terms of first- and second-order moments,
	\begin{equation}
		\braket{\hat{J}_{j}\hat{J}_{k}\hat{J}_{\ell}} \approx 2\braket{\hat{J}_{j}}\braket{\hat{J}_{k}}\braket{\hat{J}_{\ell}} - \braket{\hat{J}_{j}}\braket{\hat{J}_{k}\hat{J}_\ell} - \braket{\hat{J}_{k}}\braket{\hat{J}_{j}\hat{J}_{\ell}} - \braket{\hat{J}_\ell}\braket{\hat{J}_{j}\hat{J}_{k}}.
	\end{equation}
	This closes system (13)--(22), which was used to compute $\braket{\hat{J}_{x}(t)}$ in the second-order cumulant expansion (Gaussian approximation) of Fig.~3(a) in the main text.
	
	\section{Total Degree of Correlation At Order $n$}
	
	The measure $\mathcal{C}_{3}$ of the overall level of correlation at third-order, among the spin operators $\{\hat{J}_{x},\hat{J}_{y},\hat{J}_{z}\}$ can be generalized to correlations at any order, which could be used to quantify correlations beyond $n=3$. For a set of $r$ unique operators $\{\hat{A}_{1},\hat{A}_{2},\cdots,\hat{A}_{r}\}$ with multiplicities $k_{1}$, $k_{2}$, etc., the total degree of correlation at order $n$ is
	\begin{equation}
		\mathcal{C}_{n}(t) = \sum_{k_1}\sum_{k_2}\cdots\sum_{k_r} \bigg|n!\bigg\langle \mathcal{S}\prod_{i=1}^{r}\hat{A}_{i}^{k_{i}}  \bigg\rangle_{c} \bigg|, \text{ where } \sum_{i=1}^{r}k_{i} = n.
	\end{equation}
	Here $\mathcal{S}$ is the symmetrization (super)operator
	\begin{equation}
		\mathcal{S} = \frac{1}{n!}\sum_{\pi}\hat{\mathscr{P}}_\pi,
	\end{equation}
	where $\hat{\mathscr{P}}_\pi$ is the permutation (super)operator for the permutation $\pi$, which acts as
	\begin{equation}
		\hat{\mathscr{P}}_\pi (\hat{A}_{1} \hat{A}_{2}\cdots \hat{A}_{n}) = \hat{A}_{\pi(1)}\hat{A}_{\pi(2)}\cdots \hat{A}_{\pi(n)}.
	\end{equation}
	\section{Husimi Spin $Q$-Function}
	
	The spin $Q$-function we used is defined analogously to the Husimi $Q$-function for bosonic fields, as a trace of the density matrix over the space of coherent states. Here, the analogous basis states are the spin coherent states \cite{radcliffe_some_1971}
	\begin{equation}
		\ket{\theta,\phi} = \ket{\zeta} = \frac{1}{\sqrt{(1 + |\zeta|^2)^N}} \sum_{p=0}^{N}\sqrt{\frac{N!}{p!(N-p)!}}\ \zeta^p \ket{J,J-p},
	\end{equation}
	where $\zeta = \tan(\theta/2)e^{i\phi}\in\mathbb{C}$ and $\ket{J,J-p}$ is a Dicke state (eigenstate of $\hat{J}_{z}$) with $J=N/2$. The spin $Q$-function is then
	\begin{equation}
		Q(t,\theta,\phi) = \braket{\theta,\phi|\hat{\rho}(t)|\theta,\phi}.
	\end{equation}
	Which allows us to compute, via a separate computation of the density operator $\hat{\rho}(t)$, the visualizations of quantum states in Figs.~1,2,3, and 5.
	
	\section{Details on Quench Protocol}
	
	Our quench protocol is a piecewise linear ramp from a dispersive stage of a superradiant pulse with $\omega = \omega_\text{max}$ and $\lambda = \lambda_\text{max}$ to a dissipative stage with $\omega = \omega_\text{min}$ and $\lambda = \lambda_\text{min}$ over a ramp time $T_\text{ramp}$; here the ramp finishes at time $T_\text{pulse}$, when the pulse has its mean spin direction close to the equator of the Bloch sphere.
	
	Formally, the protocol is given by
	\begin{align}
		\omega(t) &= \begin{cases}
			\omega_\text{min},\ &\text{if}\  t < T_\text{pulse} - T_\text{ramp}, \\
			\omega_\text{max} + \dfrac{(\omega_\text{min} - \omega_{\text{max}})t}{T_\text{ramp}} - \dfrac{(T_\text{pulse} - T_\text{ramp})(\omega_\text{min} - \omega_{\text{max}})}{T_\text{ramp}}, \ &\text{if} \ T_{\text{pulse}} - T_\text{ramp} \leq t \leq T_\text{pulse}, \\
			\omega_\text{min}, \ &\text{if} \ t > T_\text{pulse},
		\end{cases} \\
		\lambda(t) &= \begin{cases}
			\lambda_{\text{max}}, \  &\text{if}\ t < T_\text{pulse} - T_\text{ramp}, \\
			\lambda_\text{max} + \dfrac{(\lambda_\text{min} - \lambda_\text{max})t}{T_\text{ramp}} - \dfrac{(T_\text{pulse} - T_\text{ramp})(\lambda_{\text{min}} - \lambda_\text{max})}{T_\text{ramp}}, \ &\text{if}\ T_\text{pulse} - T_\text{ramp} \leq t \leq T_\text{pulse}, \\
			\lambda_{\text{min}}, \ &\text{if} \ t > T_\text{pulse}.
		\end{cases}
	\end{align}
	It is illustrated in the inset panel of Fig.~5(a), where it was used to compute the third-order cumulants and, hence, $\mathcal{C}_{3}(t)$.

	\bibliography{DispFluctRefs.bib}